\documentclass[12pt]{iopart}

\usepackage{graphicx,epstopdf}
\usepackage{amssymb}
\usepackage{minibox}
\usepackage{enumitem}
\usepackage{xcolor}
\usepackage{hyphenat}

\begin{document}

\title[Cryogenic Trapped-Ion System for Large Scale Quantum Simulation]{Cryogenic Trapped-Ion System for Large Scale Quantum Simulation}


\author{G. Pagano$^1$, P.W. Hess$^2$, H. B. Kaplan$^1$, W. L. Tan$^1$, P. Richerme$^3$, P. Becker$^1$, A.~Kyprianidis$^1$,  J. Zhang$^1$, E. Birckelbaw$^1$, M. R. Hernandez$^1$, Y. Wu$^4$ and C. Monroe$^{1,5}$}

\address{
$^1$Joint Quantum Institute, Department of Physics and Joint Center for Quantum Information and Computer Science, University of Maryland, College Park, MD 20742}
\address{$^2$ Middlebury College Department of Physics, Middlebury, VT  05753}
\address{$^3$ Department of Physics, Indiana University, Bloomington, IN  47405}
\address{$^4$ Department of Physics, University of Michigan, Ann Arbor, MI  48109}
\address{$^5$ IonQ Inc., College Park, MD  20740}

\begin{abstract}
We present a cryogenic ion trapping system designed for large scale quantum simulation of spin models. 
Our apparatus is based on a segmented-blade ion trap enclosed in a 4 K cryostat, which enables us to routinely trap over 100 $^{171}$Yb$^+$ ions in a linear configuration for hours due to a low background gas pressure from differential cryo-pumping. 
We characterize the cryogenic vacuum by using trapped ion crystals as a pressure gauge, measuring both inelastic and elastic collision rates with the molecular background gas. We demonstrate nearly equidistant ion spacing for chains of up to 44 ions using anharmonic axial potentials. This reliable production and lifetime enhancement of large linear ion chains will enable quantum simulation of spin models that are intractable with classical computer modelling.

\end{abstract}


\maketitle

\section{Introduction}
Atomic ions confined in radio-frequency (rf) Paul traps are the leading platform for quantum simulation of long-range interacting spin models \cite{porras04,Kim_2009,Bohnet15, Hess20170107, Jurcevic2017}. As these systems become larger, classical simulation methods become incapable of modelling the exponentially growing Hilbert space, necessitating quantum simulation for precise predictions. Currently room temperature experiments at typical ultra-high-vacuum (UHV) pressures ($10^{-11}~\mbox{Torr}$) are limited to about 50 ions due to collisions with background gas that regularly destroy the ion crystal \cite{Zhang2017}. The background pressure achievable in UHV vacuum chambers is ultimately limited by degassing of inner surfaces of the apparatus. However, cooling down the system to cryogenic temperatures turns the inner surfaces into getters that trap most of the residual background gases. This technique, called cryo-pumping, has led to the lowest level of vacuum ever observed ($<5\cdot10^{-17}$ Torr) \cite{Gabrielse1999}. 

Here we report an experimental setup consisting of a macroscopic segmented blade trap into a cryogenic vacuum apparatus that allows for the trapping and storage of large chains of ions. 
In section \ref{setup} we describe the system design, focusing on the cryostat, the helical resonator supplying the radio-frequency (rf) drive to the ion trap, and the atomic source.
In section \ref{sec_vibrations} we report on the performance of the vibration isolation system and the improvements to the mechanical stability of the whole structure inside the 4 K region.
In section \ref{vacuum} we show pressure measurements using the ion crystals themselves as a pressure gauge below $10^{-11}~\mbox{Torr}$. We characterize the dependence of the pressure on the cryostat temperature by measuring the inelastic and elastic collision rates of trapped ions with background H$_2$ molecules. 
In section \ref{uniform} we show the capability to shape the axial potential in order to minimize the ion spacing inhomogeneity in large ion chains, and in section \ref{sec_conclusions} we summarize perspectives for future experiments.

\begin{figure*}[t!]
\centering
\includegraphics[width=1\columnwidth]{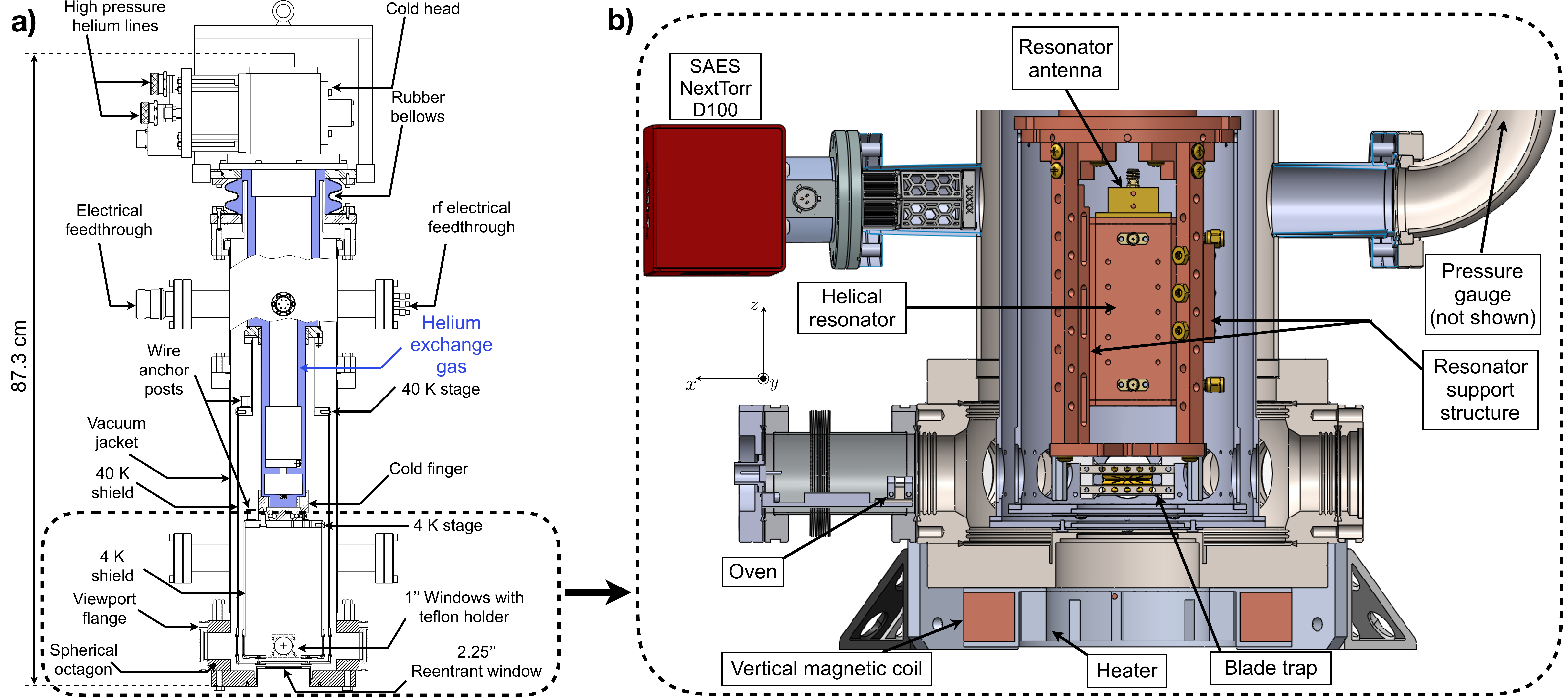}
\caption{{\bf Cryogenic vacuum apparatus.} {\bf a)} Side view section of the cryostat (courtesy of Janis Inc).  {\bf b)} Cross section view of the lower section, 90$^{\rm{o}}$ rotated with respect to a). The vertical magnetic coil is mounted on the bottom of the reentrant window flange. An aluminium fixture with heaters held on it is designed to rest in the coil's inner diameter in order to avoid water condensation on the outside face of the recessed window, when the apparatus is at 4 K.}
\label{fig_Apparatus}
\end{figure*}
\section{The Apparatus}\label{setup}

The challenge of merging atomic physics and ion trap technology with cryogenic engineering has been successfully undertaken by many groups around the world \cite{Poitzsch1996, Labaziewicz_2008, Sage2012, Schwarz2012, Niedermayr2014, Brandl2016, Alonso_2016}. Cryogenic ion traps  offer two remarkable advantages: Firstly, heating rates \cite{Turchette2000,Brownnutt2015} due to surface patch potential and electric field noise can be suppressed by orders of magnitude compared with room temperature traps, as it was first demonstrated in \cite{Labaziewicz_2008}. Secondly, the low pressures achievable via cryo-pumping  reduce the collision rate with the residual background gas, thereby enhance the lifetime of the ions in the trap.
Indeed, in standard UHV systems, the storage time of a large number of ions in a linear chain is typically limited by the probability that a ``catastrophic'' collision event causes the ion crystal to melt and the ions to be ejected from the trap.
Assuming that the collisional loss probability scales linearly with the number of ions, it is necessary to achieve a significant reduction in the background pressure in order to increase the capabilities of the trapped-ion quantum simulator platform. Recent pioneering techniques such as titanium coating and heat treatment \cite{Mamun_2014} have been shown to achieve extreme-high vacuum (XHV) in room temperature vacuum systems. However, combining room temperature XHV with an ion trap apparatus remains a challenge because many components in the vacuum system may not be XHV-compatible, such as the common electrical insulators (Kapton or PEEK). On the other hand, with a cryogenic setup, beside a lower residual gas density with respect to UHV systems, there is the additional advantage of having a lower temperature residual background gas. For this reason, the ion-molecule collisions, even when they occur, are not harmful for the ion chain. 
Moreover, a cryogenic ion-trap apparatus can offer a significant background pressure reduction at 4 K while maintaining high optical access for ion addressing and detection. Therefore, careful design is required in order to: (a) minimize room temperature blackbody radiation without limiting optical access and (b) mechanically decouple the system from the cold head vibrations while maintaining significant cooling power at 4 K. The design and the performance of the cryogenic apparatus will be described in the next subsections.

\subsection{The Cryostat}

In order to minimize the vibrations induced by the cryocooler, one possible choice is a flow cryostat \cite{Brandl2016} or a bath cryostat, which feature very low acoustic noise. However, these types of cryostats require continuous replenishment of cold liquid coolant, which is expensive and time-consuming. The alternative is to use a closed-cycle cryocooler which does not require liquid coolant to be constantly refilled and it is more convenient as it only needs external electric supply. However, a closed cycle cryocooler suffers from severe acoustic noise. To address this challenge, we decided to use a closed cycle Gifford-McMahon cryostat, whose vibrating cold finger is mechanically decoupled from the main vacuum apparatus through an exchange gas region (see Fig. \ref{fig_Apparatus}a) filled with helium gas at a pressure of 1 psi above atmosphere. The helium gas serves as a thermal link between the cold finger and the sample mount to which the ion trap apparatus is attached. A rubber bellows is the only direct mechanical coupling between the vibrating cold head and the rest of the vacuum apparatus which is sitting on the optical breadboard. This vibration isolation system allows us to keep the vibrations rms amplitude below 70 nm (see section \ref{sec_vibrations} for more details). During normal operation, the vibrating cold head is attached to the overhead equipment racks hanging from the ceiling and the vacuum apparatus is resting on an optical breadboard. However, by connecting the cold head and the cryostat mechanically, the entire system can be hoisted and moved from the optical breadboard to a freestanding structure where upgrades and repairs can be performed. 
\begin{figure}[t!]
\centering
\includegraphics[width=0.6\columnwidth]{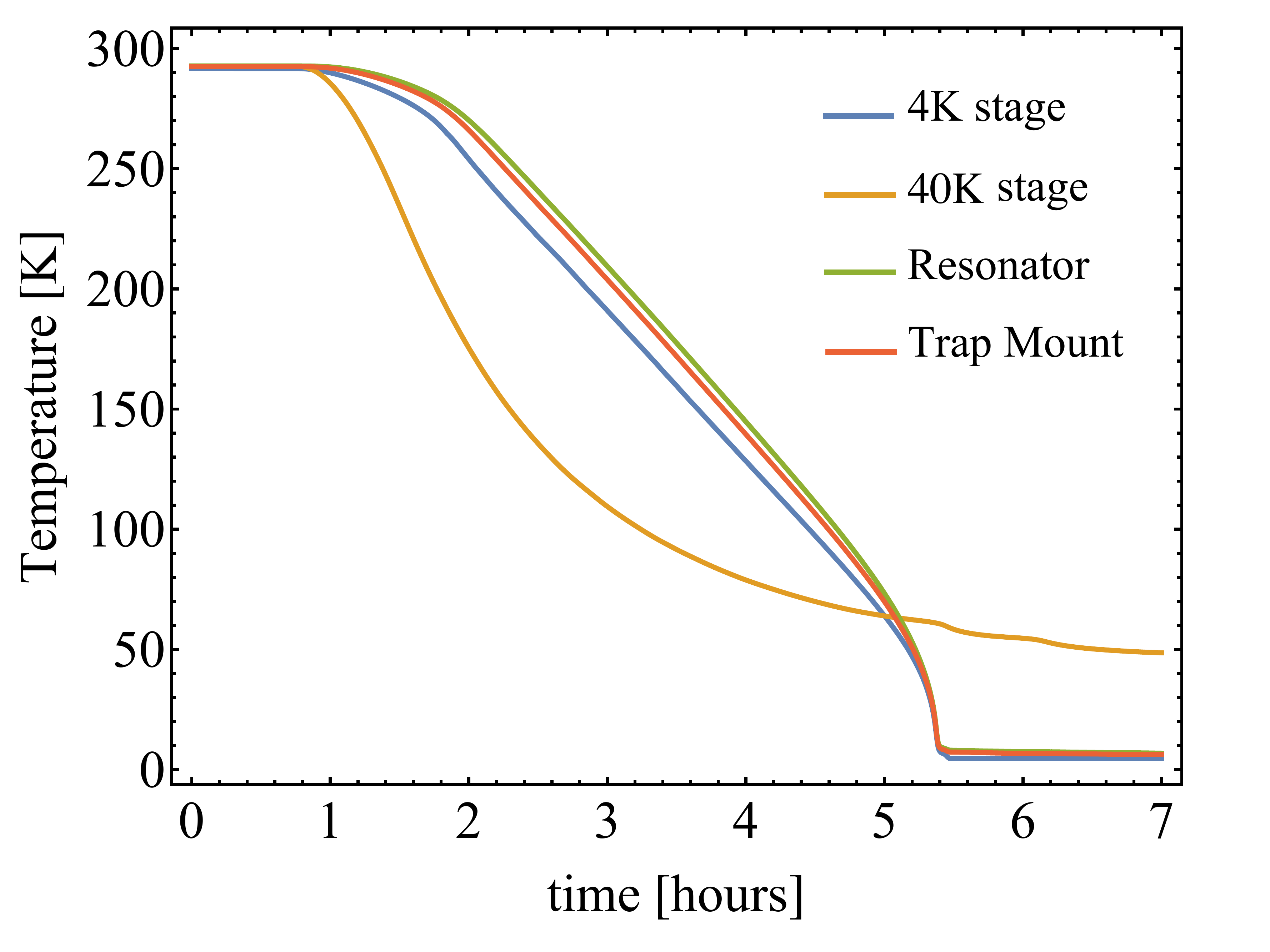}
\caption{{\bf Typical cool-down cycle}. The trap mount reaches a steady state temperature of 4.7 K while the equilibrium temperature of the helical resonator is slightly higher ($\sim$ 5.5 K) because of the reduced thermal contact with the 4 K stage. The acceleration at the end of the cool-down is due to the steep decrease in copper specific heat below 100 K \cite{EkinCryoBook}.}
\label{fig_cool-down}
\end{figure}

On top of the Cryostat (SHV-4XG-15-UHV), made by Janis Inc., sits a cold head (SRDK-415D2) that is powered by a F-70L Sumitomo helium compressor. The cold head features two stages with different cooling powers, 45~W for the 40~K stage and 1.5~W for the 4~K stage. In order to shield the ion trap apparatus from room temperature blackbody radiation (BBR), two aluminium concentric cylindrical radiation shields are in thermal contact with the two stages. Their BBR heat loads, estimated using the Stephan-Boltzmann law \footnote{The radiation heat load $\dot{Q}$ on a given surface $S$ at temperature $T_0$ from an environment at $T_1$ is given by the Stephan-Boltzmann law and it is proportional to $\dot{Q} = \sigma (T_1^4-T_0^4) /V$ where $\sigma=5.67\cdot10^{-8} $ W/(K$^4\cdot$m$^2$) and $V$ is the so called view, which is a function of the geometry, the surface $S$ and their emissivities.}, are $\dot{Q}_{40K}\sim 5.5$ W and $\dot{Q}_{4 K}\sim550$ $\mu$W, well below the cooling power of the two heat stages. 

The thermal heat load on the trap due to the electrical wiring is estimated to be negligible ($\sim100 \,\mu$W) since the temperature probes (Lakeshore DT670A1-CO), the static electrodes, and the ovens wires are all heat sunk to anchor posts (see Fig. \ref{fig_Apparatus}a) in good thermal contact with both stages.  The four SMA cables connected to the radio-frequency (rf) electrical feedthrough are not heat sunk and deliver an estimated heat load of 500 mW and 220 mW on the 40 K and 4 K stages, respectively \footnote{The heat conduction contribution of the wiring has been estimated as $\dot{Q}=A/L \int_{T_1}^{T_2}\lambda(T)dT$, where $A$ is the wire cross-section, $L$ is the wire length and $\lambda(T)$ is the temperature dependent heat conductivity and $T_1$ and $T_2$ are the temperatures of the two ends of the wire \cite{EkinCryoBook}}. In designing the apparatus, a good balance between room temperature BBR heat load and optical access has been achieved, given the total cooling power. The spherical octagon (see Fig. \ref{fig_Apparatus}) features eight 1'' diameter windows which provide optical access in the $x$-$y$ plane, held by soft teflon holders. On the bottom, the system features a 2.25'' diameter reentrant window which allows for a numerical aperture (NA) up to 0.5 to image the ions along the vertical ($z$) direction. The whole cryostat rests on an elevated breadboard to allow for ion imaging from underneath. The total heat load of the recessed window is estimated to be 2.4 W on the 40 K shield and 1.7 mW on the 4 K shield. The total heat load budget (see \ref{app_heat_loads}) is well below the total cooling power, and therefore we can cool the system down to 4.5 K in about 5 hours (see Fig. \ref{fig_cool-down}) with both the helical resonator and the ion trap inside the 4 K shield.

Before the cool down, the apparatus is pre-evacuated using a turbo-molecular pump until the pressure reading on the MKS-390511-0-YG-T gauge (not shown in Fig. \ref{fig_Apparatus}b) reaches the pressure of about ${\simeq 2\cdot10^{-5}}$~Torr. The gauge is attached through an elbow in order to avoid direct ``line of sight" exposure of its radiative heat load (1.5~W) to the 40 K shield. Since hydrogen is the least efficiently cryo-pumped gas, we added a SAES NexTorr D-100 getter and ion pump (see Fig. \ref{fig_Apparatus}b) which is attached to one of four CF40 flanges of the long bellow cross.
\begin{figure}[ht]
\centering
\includegraphics[width=0.6\columnwidth]{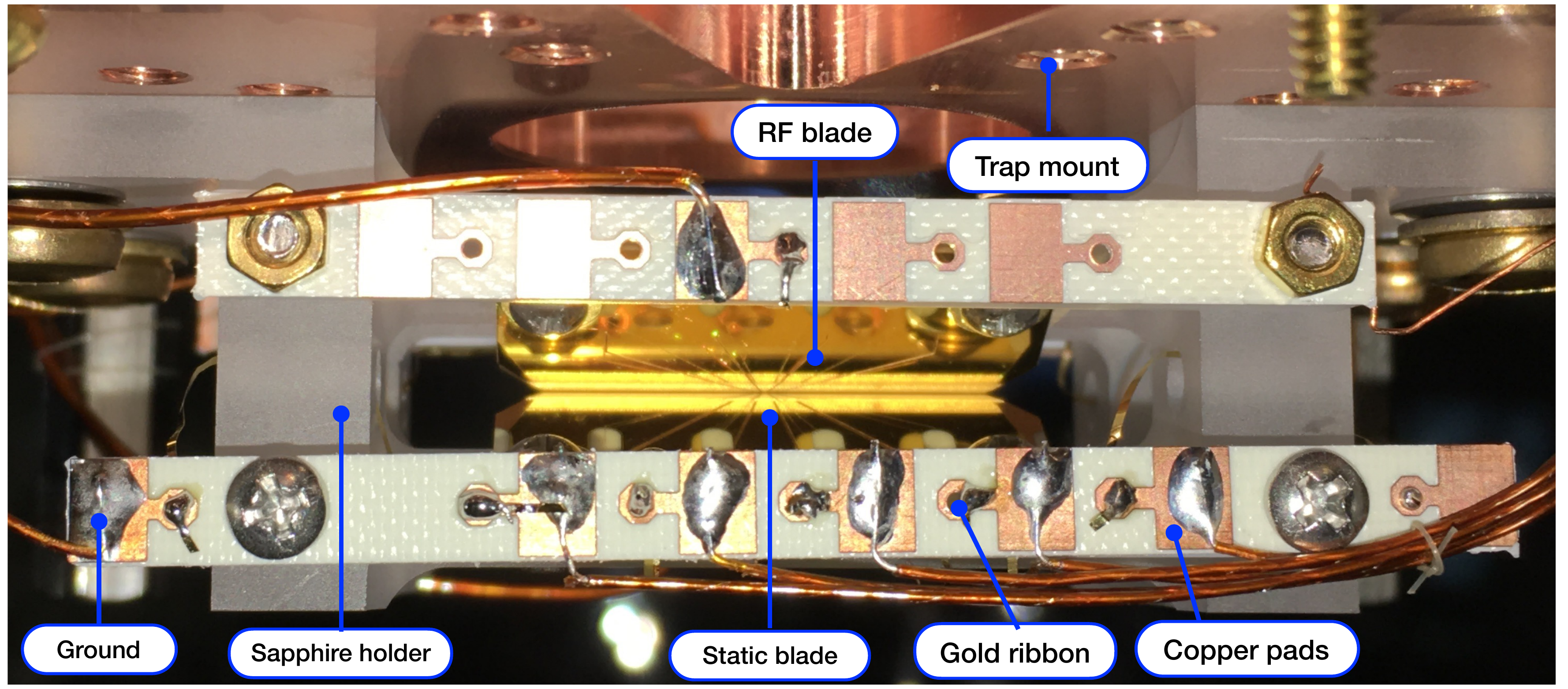}
\caption{{\bf Blade ion trap}. The blades are mounted on a sapphire holder and gold ribbons are wirebonded on top of them. The connection between the gold ribbons and the kapton wires is provided by copper pads printed on Roger 4350B PCB.}
\label{fig_trap}
\end{figure}


\subsection{Blade trap}
The blade-style ion trap was hand-assembled and aligned under a microscope with  $\simeq5\,\mu$m resolution. The blades are made of alumina and have five segments. They were cleaned with hydrofluoric acid and plasma ashed on both sides, then coated with a 100 nm titanium adhesion layer and a 1 $\mu$m gold layer at Sandia Laboratories. The coating was applied by means of multi-directional evaporation to avoid alumina exposure in the 50 $\mu$m wide gaps between the segments. The two static electrode blades are gold coated with masks so that the five segments can be biased independently, whereas the rf blades have an unsegmented coating. Gold ribbons (0.015" wide and 0.001" thick) are wire-bonded on top of the blades for electrical connections. In order to shunt rf pickup voltages on the static blades, 800 pF ceramic capacitors connect each static electrode segment to ground. The capacitors (DIGIKEY 399-11198-1-ND) are made of NP0, whose capacitance has small temperature dependence down to 4 K. We chose to solder the capacitors onto the gold ribbons as we could not find any capacitor made of low dielectric constant material that could be wirebonded directly to the gold ribbon. Standard solder is cryo-compatible \cite{EkinCryoBook}, so we used it to connect the other end of the gold ribbons to the external copper pads (see Fig. \ref{fig_trap}) for wiring, instead of using the invasive spot-welding procedure usually followed for UHV blade traps. The blades are mounted on a sapphire holder in a 60$^{\rm{o}}$/30$^{\rm{o}}$ angle configuration, which allows good optical access both in the $x$-$y$ plane and along the vertical $z$ direction, and strongly breaks rotational symmetry so that the trap principal axes are well-defined. The distances between the electrodes tips are 340~$\mu$m/140~$\mu$m and the ion-electrode distance is 180~$\mu$m. The $x$-$y$ plane features 0.1 NA through the eight 1'' windows, which are used for Doppler cooling, detection, optical pumping (369 nm), photo-ionization (399 nm), repumping (935 nm) and Raman (355 nm) laser beams \cite{Olmschenk2007}. We can perform high-resolution imaging along the vertical $z$ direction (see Fig. \ref{fig_Apparatus}b) since we have a 3.5 cm working distance on a 2'' window, allowing for an objective of up to 0.5 NA. 

In order to provide eV-deep trapping potentials and high trap frequencies, the blade trap needs to be driven with hundreds of Volts in the radio-frequency range. Based on the COMSOL simulation model of the blade trap, we need about $V_{\rm rf}=480$ V amplitude to get a $\omega_{\rm tr}= 2\pi \times 4$ MHz transverse trap frequency with a rf drive at $\Omega_{\rm rf} = 2\pi \times 24$ MHz.  Considering the calculated trap capacitance $C_t= 1.5$ pF, we can estimate the power dissipated \cite{Stick2006} on the blades to be  $P_d=\frac{1}{2} \Omega_{\rm rf}^2 C_t^2  V_{\rm rf}^2 R_t \simeq 1~\mbox{mW},$ where  
$R_t$ is the resistance of the $1 \,\mu$m gold layer on the blade, which is estimated considering that the gold skin depth is $\delta \simeq250~\mbox{nm}$  at 4 K.
Moreover, the blades are efficiently heat sunk, as they are mounted on a holder made of sapphire, which presents the double advantage of a better thermal conductivity compared to macor or alumina holders and a well matched coefficient of thermal expansion with the alumina blade substrate.
\subsection{Helical Resonator}
A helical resonator \cite{Macalpine1959} enables impedance matching between a radiofrequency source and the trap by acting as a step-up transformer. The parasitic capacitance and inductance of the 70 cm coaxial rf transmission cable between the vacuum feedthrough and ion trap would make the impedance matching of the tuned resonator and trap circuit~\cite{Siverns2012} very difficult, thereby limiting the rf voltage that could be delivered to the trap without incurring in resistive heating of the rf cables. Therefore, the apparatus is designed to host the helical resonator in the 4 K region as close as possible to the blade trap.
\begin{figure}[t!]
\centering
\includegraphics[width=0.6\columnwidth]{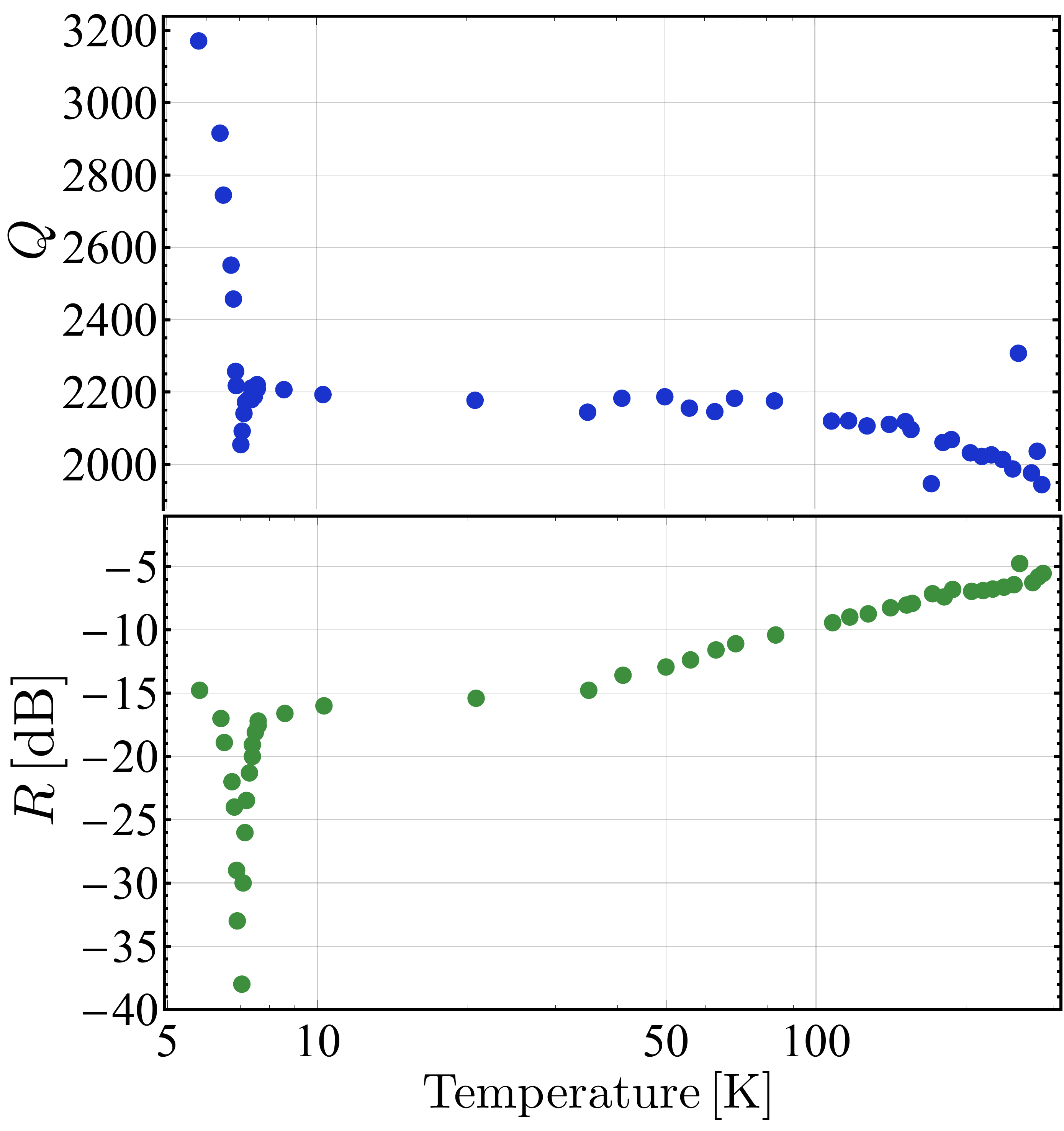}
\caption{
{\bf Resonator Q-factor and rf reflected power as a function of the temperature.} The steep variation at low temperatures is due to the sharp decrease in the copper resistivity below 100 K \cite{EkinCryoBook}, whose effect is delayed as the bifilar coil is in poor thermal contact with the copper can. The loaded quality factor $Q_{\rm{load}}$ (see text for details) increases from 210 up to 900 during the cool-down.}
\label{fig_Q}
\end{figure}

Inside the helical resonator, we have wound a bifilar coil whose two components are connected to the two rf blades. In this way, each blade can have an independent static potential offset. The two coils are held in place by teflon holders and they are shorted at rf with a 400 nF capacitor. The resonator is made of solid copper with a 2.3'' inner radius, whereas the bifilar coil features a 1.5'' radius and a 0.19'' pitch.
We inserted a capacitive 100:1 pick-off of the rf voltage inside the resonator to monitor and actively stabilize the transmitted voltage amplitude to the trap blades \cite{Johnson2016}. The self-inductance  $L_{\rm res} = 2~\mbox{$\mu$H}$ and self-capacitance $C_{\rm res} = 8~\mbox{pF}$ have been measured loading the resonator with different test-capacitors. 

At room temperature the intrinsic (``unloaded'') resonator quality factor is ${Q=1050}$, where $Q$ is defined as:
\begin{equation}
\label{eq_}
Q=\frac{2\, Q_{\rm{load}}}{1-\sqrt{R}},
\end{equation}
where $R$ is the reflected rf power due to impedance mismatch (see Fig. \ref{fig_Q}) and $Q_{\rm{load}}=\Omega_{\rm rf}/\rm{FWHM}$ is the loaded Q-factor, which takes into account the response of the helical resonator connected to the blade trap. The resonator is inductively coupled with a small 0.5'' diameter antenna-coil, whose position can be tuned to reach critical coupling. At 4 K, the resistance of the whole rf circuit is reduced and two effects take place: the $Q$ value at critical coupling increases up to 3170 and the impedance matching condition changes (see Fig. \ref{fig_Q}). The 60\% increase in the resonator quality factor is lower than what would be expected from a simple estimate based on the decrease of copper resistivity and skin depth at 4 K. This is likely explained by oxide layers on the copper surface or by the additional resistance contribution of solder connections in the resonator. In order to compensate the temperature induced resistance change, we reduced the mutual inductance between the antenna and the bifilar coil by pulling out the antenna holder (see Fig. \ref{fig_Apparatus}b) away from the optimal position at room temperature. During the cool-down, the drive frequency $\Omega_{\rm rf}/2\pi$ increases typically by $0.6\%$, which is explained by the reduction in resonator self-capacitance and self-inductance induced by thermal contraction of the copper.  

\subsection{Atomic source}
In order to load the ions in the blade trap we resistively heat up the atomic source, which is encased in a 0.5'' long stainless steel (SS) tube with a tantalum wire spot welded to each end. The oven is supported by a macor holder, made of two 0.05'' ID halves enclosing the SS tube to ensure stable pointing and to provide thermal insulation. We noted that having the atomic sources in the 4 K region, as close as possible to the trap, causes numerous problems: the heat load coming from the ovens was enough to make the whole copper structure expand more than the trap clearance, requiring the Doppler cooling beam to be steered accordingly for trapping. Moreover, the whole ramp up of the ovens had to be done gradually as the thermal gradients could break the tantalum wires, resulting in a total average loading time of 1h and 30 min. For all these reasons, we decided to move the atomic source to the room temperature sector of the vacuum apparatus (see Fig. \ref{fig_Apparatus}b). 
\begin{figure*}[t!]
\centering
\includegraphics[width=1\columnwidth]{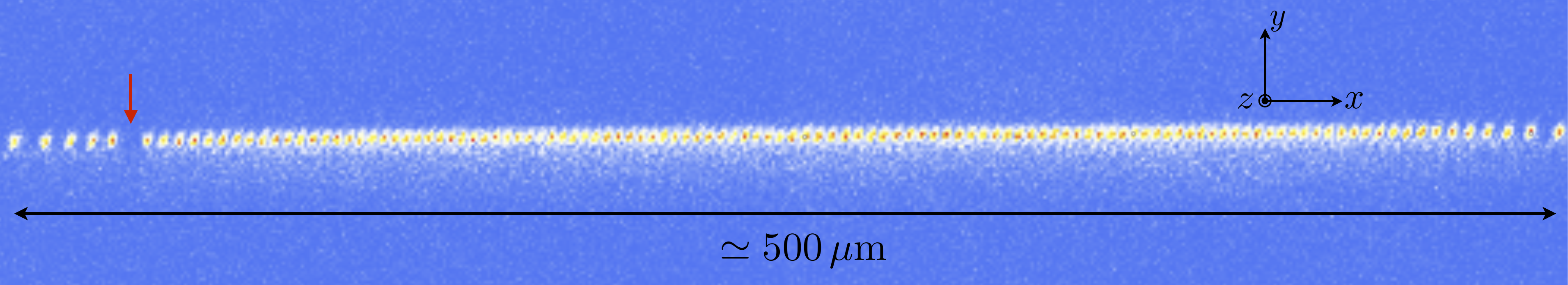}
\caption{{\bf Linear chain of 121 $^{171}$Yb$^{+}$ ions}. In this case $\omega_{y}/2\pi=1.5~\mbox{MHz}$ and $\omega_x/2\pi=35~\mbox{kHz}$. The axial confinement has been relaxed to resolve all the center ions with a 0.13 NA objective. In order to fit the whole chain in the CCD camera, we took two images of the left and right part of the chain by moving the objective along the trap axis. There is only one dark ion in the $^{2}F_{7/2}$ state on the left part of the chain, indicated by a red arrow.}
\label{fig_ions}
\end{figure*}
The oven is in a 2.93'' long UHV bellow so that under vacuum the atomic source is 11.9 cm away from the trap. The atomic beam is collimated by two pinholes mounted on the 4 K and 40 K shields and is aligned on the trap axis $(x)$ to minimize the exposure of the electrode gaps. This prevents shorts or the formation of ytterbium oxide layers on the blades when the system is vented. The oven's macor holder sits on an aluminium bar which is screwed down to the feedthrough of the bellows and whose height is designed to match the trap axis location when the apparatus is cold (see Fig. \ref{fig_Apparatus}b). The use of the bellows was originally intended to enable one to steer the atomic beam to maximize the loading rate, but the alignment by design was enough to provide a satisfactory loading rate. The generation of the atomic flux is achieved by the Joule effect, dissipating 1.5 W on the $0.22$ Ohm oven resistance. With this setting, in approximately 4 minutes (including oven warm-up time) we can load about 50 ions by using 2 mW of 399 nm and 350 mW of 355 nm laser light to perform a two-step photoionization process.
\section{Vibration isolation system}\label{sec_vibrations}
\begin{figure}[t!]
\centering
\includegraphics[width=0.6\columnwidth]{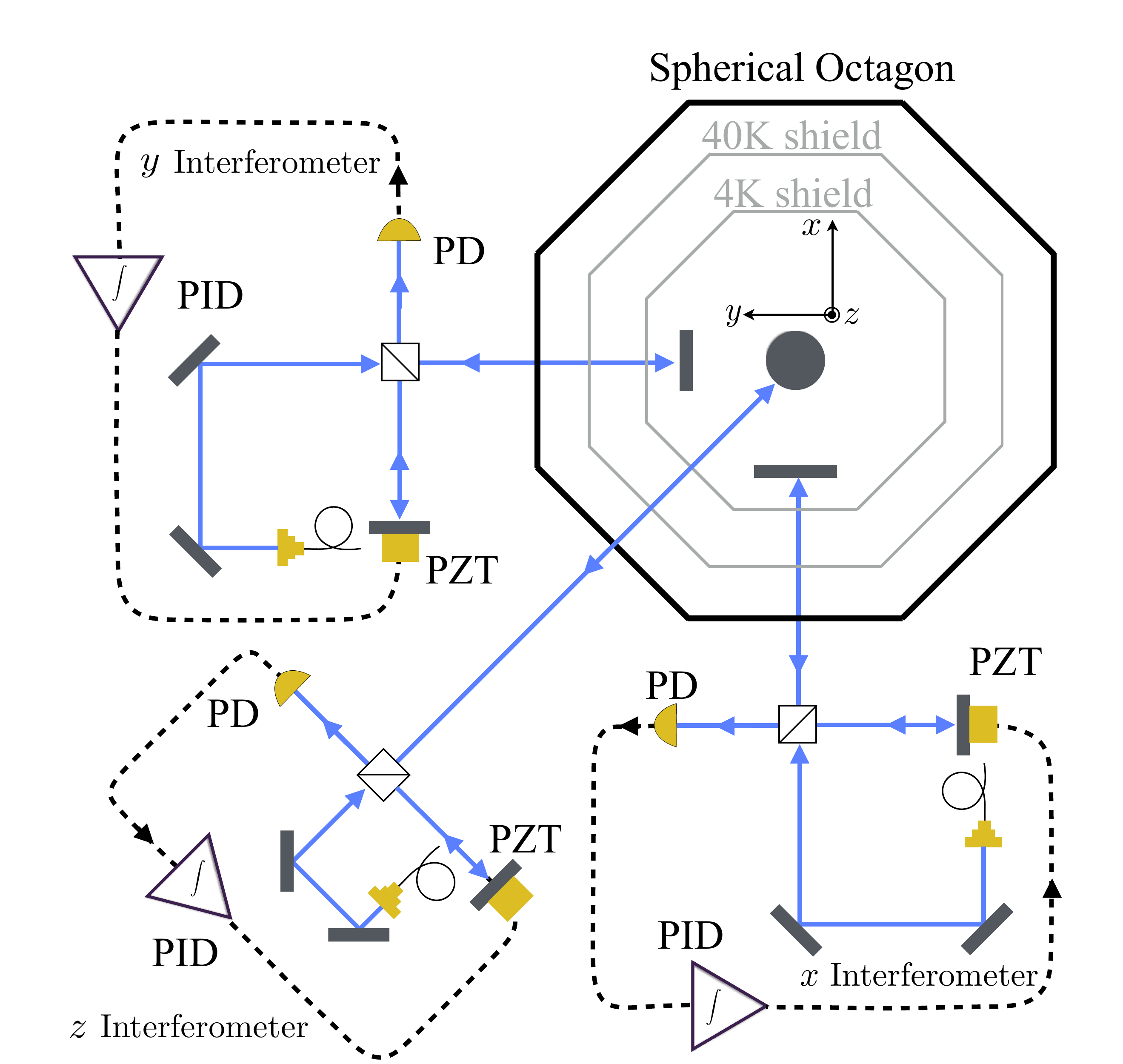}
\caption{{\bf Interferometric setup for vibration measurements}:  Three mirrors are mounted inside the cryostat and attached to the trap mount along the three principal axes of the trap. Three different fibers deliver laser light to three interferometers with piezo-mounted mirrors (PZT) which lock the photodiode signal to a fringe with a feedback loop. The PID voltages output to the piezos compensate for vibrations and are used to monitor their amplitude and frequency.}
\label{fig_interferometers}
\end{figure}
Since the Gifford-McMahon cold head compression and expansion cycles produce a significant amount of acoustic vibrations, it is crucial to assess and improve the performance of the vibration isolation system (VIS). This system consists of an exchange gas region filled with helium gas, to provide a thermal link between the vibrating cold head and the cold finger. A rubber bellows (see Fig. \ref{fig_Apparatus}a) is the only mechanical connection between the cold head and the lower part of the apparatus and acts as a vibration damper.

We are interested in both the timescale and the amplitude of the residual acoustic vibrations. Indeed we plan to perform quantum simulation experiments lasting up to 10 ms, creating spin-spin interactions \cite{Kim_2009} through stimulated two-photon Raman processes driven by a pulsed laser at $\lambda = $ 355 nm. Therefore, any ion chain displacement of the order of $\lambda$ during the laser interaction time would result in an unwanted phase shift experienced by the ions. 

\begin{figure*}[t!]
\centering
\includegraphics[width=1\columnwidth]{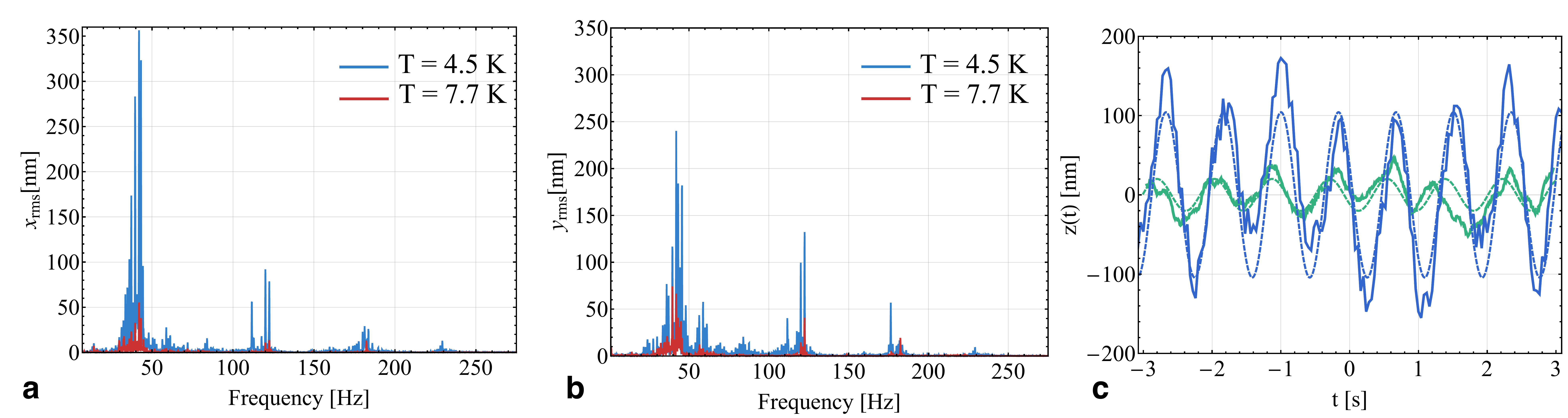}
\caption{{\bf Vibrations along the three trap principal axes}:  a-b) In-plane $x$-$y$ vibrations for trap mount temperatures $T=4.5$ K and $T=7.7$ K. Rms amplitudes along $x$($y$) are reduced by a factor of 5(6) respectively by raising the temperature of the cryostat above the helium boiling point. The RBW is 0.1 Hz. c) Vertical vibration induced by the cryostat fitted with a sine at 1.2 Hz frequency (dashed lines). The blue (green) line refers to the vibrations before (after) improving the static support of the breadboard, yielding a 40 nm pk-pk oscillations.}
\label{fig_Vibration_vs_T}
\end{figure*}

We characterized the mechanical stability of the whole apparatus by interferometric measurements. We removed the trap and placed three mirrors on the trap mount, along the three principal axes: the Raman direction $y$, the trap axis $x$ and the vertical imaging direction $z$ (see Fig. \ref{fig_interferometers}). In this way, every movement of the trap region with respect to the breadboard and the table can be measured with a few nanometers resolution. In order to reliably assess the displacement of the trap by counting the number of fringes induced by the acoustic vibration, three piezo-mounted mirrors are used on the reference path of the Michelson interferometers to lock the interferometer to a fringe (see Fig. \ref{fig_interferometers}). The servo loops have 1.8 kHz bandwidth and therefore they can fully compensate acoustic noise over the frequency range of interest (up to 300 Hz). The voltage-to-distance conversion of the piezo-mounted mirrors has been characterized on a static interferometer: in this way, the output voltages applied to the piezos are a direct measurement of the displacement of the trap mount along the three principal axes of the trap.

\begin{figure}[t]
\centering
\includegraphics[width=0.6\columnwidth]{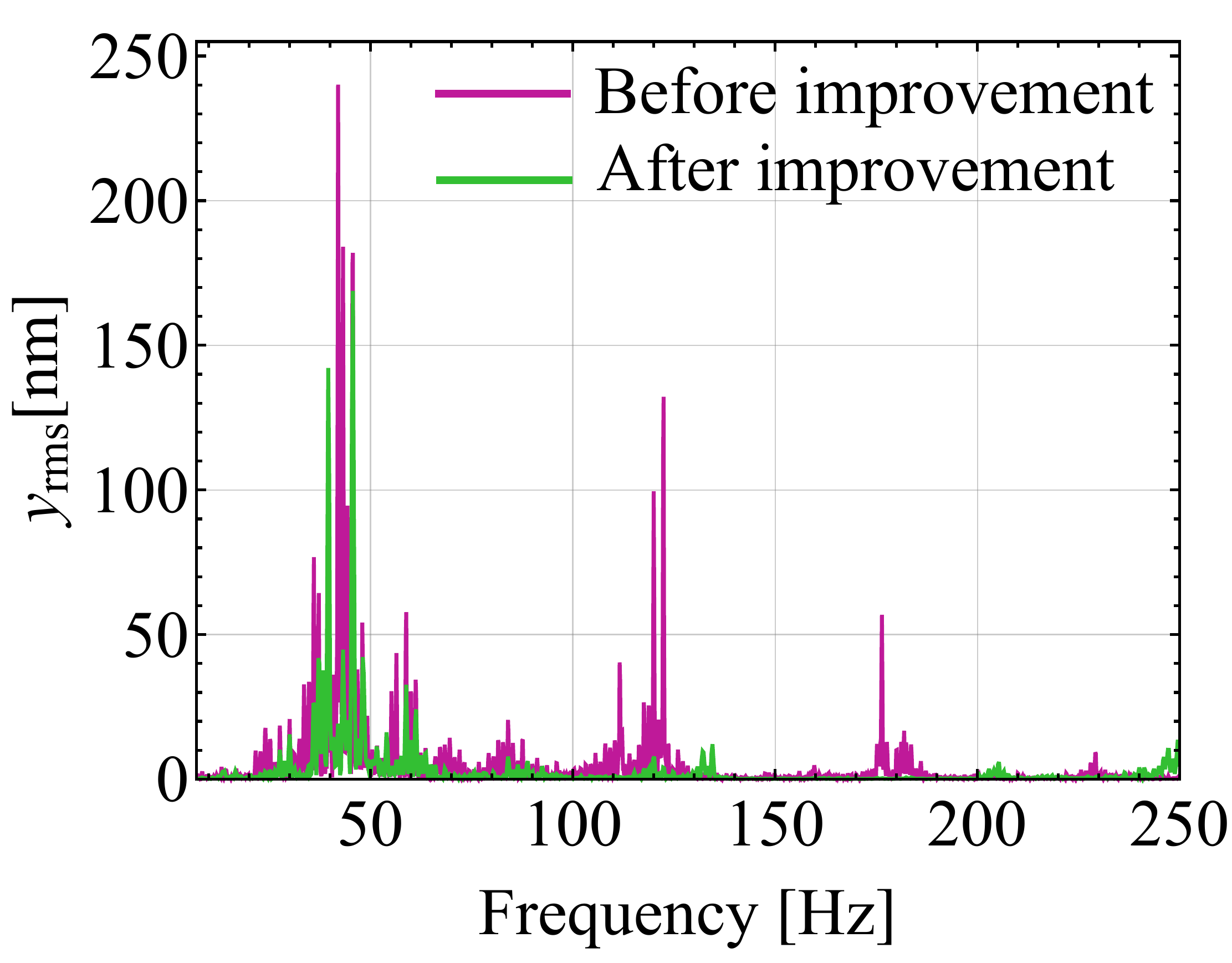}
\caption{{\bf Improved vibrations along the $y$ axis}: After improving the resonator static support structure, the frequency modes above 100 Hz shown in Fig. \ref{fig_Vibration_vs_T}b are suppressed by a factor of 10. The dominant acoustic modes remaining after the improvement are at 39 and 45 Hz. For both datasets, $T=4.5~\mbox{K}$. The RBW is 0.1 Hz. }
\label{fig_VibrationY}
\end{figure}
In the $x$-$y$ plane, we observe that the main contribution is given by a peak around 40 Hz, which is attributed to the normal mode of the cryostat 60 cm long lever arm. Finite element analysis suggests that the vibration modes at higher frequencies (120 Hz and 282 Hz peaks) are to be attributed to the mechanical structure inside the 4 K shield, made up of the resonator and the trap mount. We noticed that most of these vibrations were driven by the vibrating cold head through liquid helium condensed at the bottom of the exchange gas region at the lowest operating temperature  (see Fig. \ref{fig_Apparatus}b). This condensation results in a reduced performance of the VIS, which can be overcome by heating up the 4 K stage and operating above the He boiling point. With the 4 K stage at $T = 7.7$ K, the rms displacements in both the $x$ and $y$ directions are reduced by a factor of 5, as shown in Fig. \ref{fig_Vibration_vs_T}. 
The vibration modes above 100 Hz are the most problematic ones as they are on a similar timescale of a typical quantum simulation experiment. In order to eliminate these higher frequency modes, we strengthened the resonator static support, making the whole 4 K structure stiffer. By doing so, we were able to suppress the higher frequency modes by more than an order of magnitude (see Fig. \ref{fig_VibrationY}). This leaves only two very well defined normal modes at 39 Hz and 45 Hz, which have been also suppressed with respect to Fig. \ref{fig_Vibration_vs_T}b. 
These lower frequency vibration modes constitute a manageable problem for our typical experiment duration as we could apply spin-echo schemes to compensate the unwanted 40 Hz phase variation. In addition, we designed a permanent mirror holder compatible with the trap in order to monitor in real time the vibrations along the Raman $y$ direction. In this way it will be possible to compensate the unwanted phase shift via a feed-forward to the AOM phase or to an EOM in the Raman path.

We also investigated the mechanical stability along the vertical $z$ direction, finding a very well defined oscillation at the cold head vibration frequency (1.2 Hz) with a peak-to-peak amplitude of 200 nm, as shown in Fig. \ref{fig_Vibration_vs_T}c. We ascribed this almost entirely to a vibrational mode of the whole breadboard on which the cryostat is sitting. Indeed, we observe exactly the same oscillation using a reference mirror attached underneath the breadboard, instead of the mirror attached to the trap mount inside the cryostat. In order to reduce these slow oscillations, we added more static support to the elevated breadboard and thus reduced the peak-to-peak amplitude to about 40 nm.


\section{Characterization of Cryogenic Vacuum} \label{vacuum}
The background pressure requirements for ion trap experiments are demanding primarily for two reasons. First, the ions interact with the residual neutral molecule gas with the long range $\sim r^{-4}$ potential, which increases the collision rate compared to e.g. neutral-neutral collisions rates, governed by Van-der-Waals $\sim r^{-6}$ potentials. Secondly, since the Paul trap is not static, the collisions with a bath of neutral particles can induce heating by displacing the ions diabatically with respect to typical rf timescales, depending on the mass imbalance and on the instantaneous rf phase at which the collision occurs \cite{Zipkes2011, Cetina_2012, Chen_2014}. As a consequence of this instantaneous and random amplification of the ion motion, the ion crystal melts, avalanche rf heating takes place, and the ions are either ejected out of the trap or left in highly excited orbits where laser cooling is inefficient. The ion chain lifetime depends strongly on the Mathieu parameter $q\simeq 2\sqrt{2}\omega_{\rm tr}/\Omega_{\rm rf}$ of the trap in consideration: once one ion is displaced by a collision, nonlinearity coming from the Coulomb repulsion will cause the ion's kinetic energy to grow at a rate that scales as a power law of $q$ with an exponent greater than or equal to 4 \cite{Ryjkov2005}.
The $q$ parameter sensitivity is also related to higher order terms in the rf trapping potential expansion beyond the first order quadrupolar contribution. These terms are associated with nonlinear resonances inside the stability region of a linear Paul trap \cite{Alheit1996,Drakoudis2006} that are likely to accelerate the ion loss once the crystal has melted.

We can qualitatively estimate, from kinematic considerations, the energy acquired by an ion $i$ at rest after an elastic collision with a background molecule with incoming energy $E_m$:
\begin{equation}
\label{eq:}
\Delta E_i =\frac{4\xi}{(1+\xi)^2} \sin^2(\theta_{\rm sc} /2)E_m,
\end{equation}
where $\xi=M_m/M_i$ is the mass-imbalance parameter and $\theta_{\rm sc}$ is the scattering angle
 (for details, see \ref{app:collision}). In both a room-temperature UHV system and a cryogenic apparatus, the residual background gas is made mostly of hydrogen molecules (H$_2$), the least efficiently cryo-pumped gas after helium, leading to $\xi=0.011$ for $^{171}$Yb$^+$ ions. Considering the thermal energy of H$_2$ molecules $\langle E_{\mbox{\tiny{H}}_2}\rangle = 3/2 \, k_B \times 4.5$ K, we obtain, averaging over impact parameters, a mean ion energy increase of $\langle\Delta E_{\mbox{\tiny{Yb$^+$}}} \rangle \simeq k_B\times 150~\mbox{mK}$ per elastic collision.

In this cryogenic vacuum system, we have not observed a catastrophic ion loss induced by collisions, even for chains of above 100 ions at $q=0.35$. In the room temperature UHV experiment, with same $q$, the lifetime of a chain of about 50 ions is on average 5 minutes \cite{Zhang2017}.
We can attribute the enhanced lifetime to two factors: on one hand differential cryo-pumping allows us to reduce the residual gas density with respect to standard UHV systems, therefore reducing the overall collision rate. Secondly, even when a collision occurs, the average energy transfer $\langle\Delta E_{\mbox{\tiny{Yb$^{+}$}}} \rangle$ is about 60 times lower than in a room temperature UHV experiment and the crystal melting is less likely to happen as the Doppler cooling laser can efficiently recapture and recrystalize the ions. We notice that in a UHV apparatus $\langle\Delta E_{\mbox{\tiny{Yb$^{+}$}}} \rangle\simeq k_B \times10~\mbox{K}$ for H$_2$ molecules, which, according to our numerical simulations, is not sufficient to displace the ions enough to melt the crystal. According to numerical simulations (see \ref{app_numerical_simulation}), catastrophic collisions in a room temperature UHV system are most likely caused by rare collisions with heavier residual background gases (N$_2$, CO$_2$ and H$_2$O), that in a cryogenic system are completely frozen.
\begin{figure}[t]
\centering
\includegraphics[width=0.6\columnwidth]{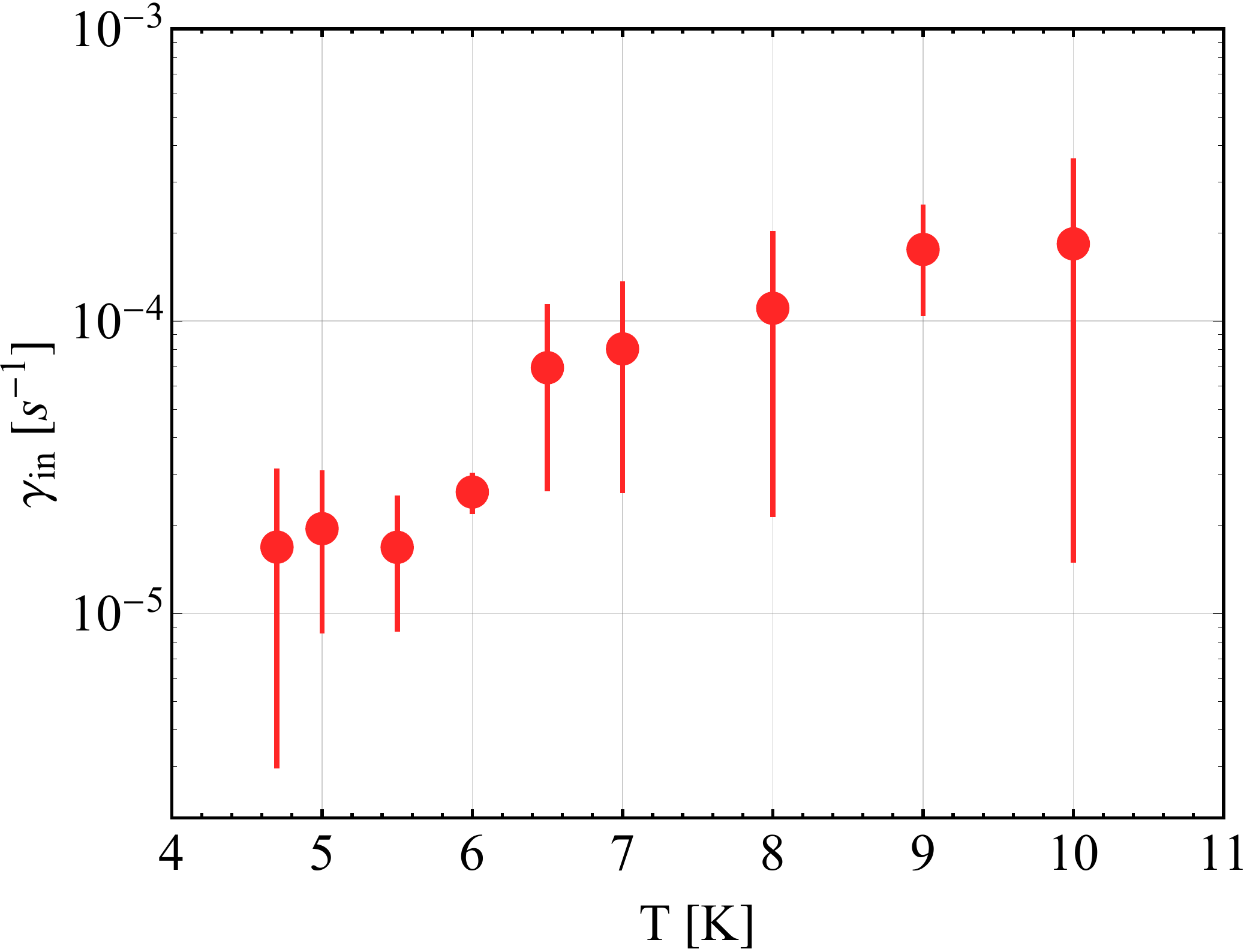}
\caption{{\bf Average dark ion rate as a function of temperature}: the data have been acquired with 33 ions for a time period varying from 12 hours to 3 hours. The reported inelastic rate is per ion, estimated by averaging the time intervals between dark events. The error bars are the time interval standard deviations of each data set. The size of the error bars is caused by the poor statistics, due to the rare occurrence of inelastic collision processes.}
\label{fig_inelastic_rate}
\end{figure}

In order to quantify the background pressure in the vacuum system, we could not use a hot cathode ionization gauge as it can generate additional gas load. Moreover, the gauge in use measures the pressure of the room temperature vacuum region (see Fig. \ref{fig_Apparatus}b), which is on the order of low $10^{-9}$ Torr when the apparatus is cold. Since both the 40 K and the 4 K cryogenic regions are not vacuum sealed, it is difficult to estimate the differential cryo-pumping between the room temperature region and the inner part of the apparatus. For this reason, we used the ion crystal as a pressure gauge by measuring collision rates with the molecular hydrogen background gas as a function of the cryostat temperature. 

The ion-neutral molecule interaction is described by a  $\sim r^{-4}$ potential, stemming from the interaction between the
charge $e$ of the ion and the induced electric dipole moment of the molecule with static polarizability $\alpha$:
\begin{equation}
\label{eq_Pot}
U(r)=-\frac{\alpha}{2}\frac{e^2}{4\pi\varepsilon_0 r^4}=-\frac{C_4}{r^4}.
\end{equation}
We can safely assume the validity of the classical Langevin model \cite{Langevin1905} with no quantum corrections, since the average energy of the incoming H$_2$ molecule $\langle E_{\rm H_2}\rangle$ is much larger than the $p$-wave centrifugal barrier associated with the potential (\ref{eq_Pot}), which can be approximated as $E_4=\hbar^2/2\mu R_4^2 \simeq k_B \times3\mbox{ mK}$, with $R_4=(2 \mu C_4/\hbar^2)^{1/2}$ and $\mu$ the reduced mass \cite{Jachymski2013}.

In the Langevin model  the collision rate $\gamma$ is independent of the energy of the incoming particle and it is directly proportional to the density $n$, or equivalently to the pressure $P$ assuming the ideal gas law. This results in:
\begin{equation}
\label{eq:Pressure}
\gamma=n \,e \sqrt{\frac{\alpha_{H_2}\pi}{\mu\,\varepsilon_0}},\,\,\,\, n=\frac{2}{3} \frac{P}{k_B T}
\end{equation}
\begin{figure}[t]
\centering
\includegraphics[width=0.95\columnwidth]{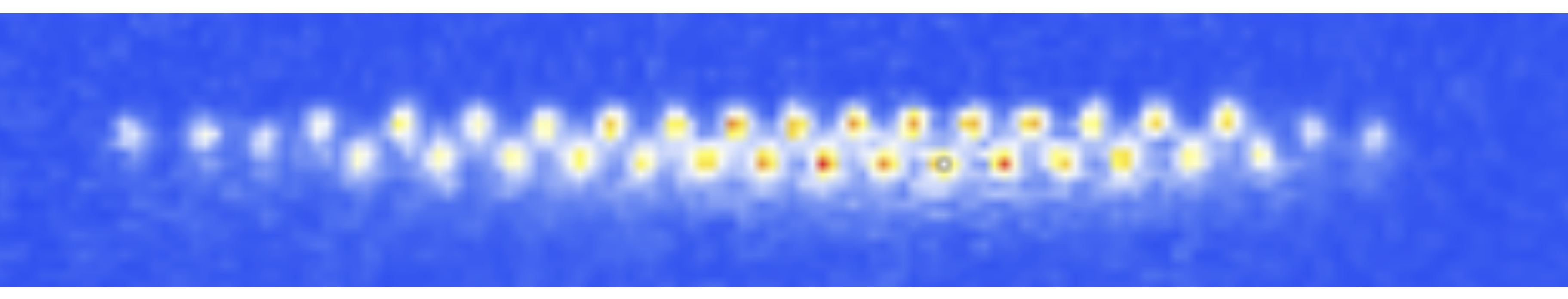}
\caption{{\bf Zig-Zag ion chain}: Zig-zag chain with $N= 35$ ions with $(\omega_x,\omega_y,\omega_z)=2\pi\times(67,613,632)~\mbox{kHz}$.}
\label{fig_elastic_rate}
\end{figure}
In order to estimate the residual background pressure inside the cryostat, we measured the rate at which dark ions are produced. Indeed, whenever a collision occurs, there is a finite unknown probability $P_{\rm in}$ that two inelastic processes take place: (a) optically excited ions $^{171}$Yb$^{*+}$ in the $^{2}P_{1/2}$ or $^{2}D_{3/2}$ states, both populated by the Doppler cooling light at 369 nm, are subjected to collisional quenching that leads to population trapping in the metastable ${^2}F_{7/2}$ state \cite{Lehmitz1989}. (b) Molecule association: an optically excited ion has enough energy to chemically react with H$_2$, breaking its bond and forming an ytterbium hydride (YbH$^+$) molecule \cite{Sugiyama1997}. In both cases the ions stop scattering Doppler cooling photons and appear as missing in the ion chain as imaged on an EMCCD camera (Andor iXon 897).
Therefore, we can extract a relative measurement of the pressure by recording the occurrence rate of dark ions, namely $\gamma_{\rm in} = P_{\rm in}\gamma$. We measured the dark ion rate as a function of the temperature of the cryostat (see Fig. \ref{fig_inelastic_rate}) observing an increase by an order of magnitude, with a temperature increase of 5.5 K. By comparing the dark ion rate in the cryogenic vacuum system with the room temperature UHV system with a gauge-measured pressure of $1\cdot10^{-11}$ Torr and $\gamma_{\rm in} = 2\cdot10^{-4}~\mbox{s$^{-1}$}$ per ion, we can infer the residual background pressure as:
\begin{equation}
\label{eq_}
P_{4\rm K}=P_{300{\rm K}}\frac{\gamma^{(4{\rm K})}_{\rm in} }{\gamma^{(300{\rm K})}_{\rm in} }\frac{k_B T_{\rm 4 K}}{k_B T_{\rm 300 K}}.
\end{equation}
With this method, we estimate $P_{4\rm K}<10^{-13}~\mbox{Torr}$.

\begin{figure*}[t!]
\centering
\includegraphics[width=1\columnwidth]{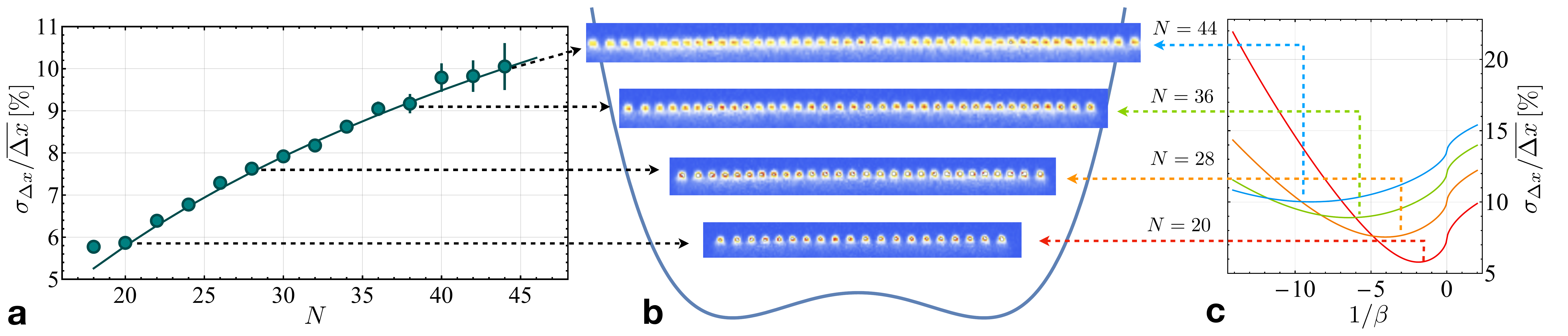}
\caption{{\bf Uniformly spaced ion chains}. 
a) The spacing variance to mean ratio $\Delta \sigma_x/\overline{\Delta x}$ as a function of the number of ions in the linear chain. Data points are taken in the optimized electrode configuration resulting in a minimal inhomogeneity, which corresponds to the theoretical prediction for a quartic potential. The spacings value have been chosen according to practical considerations on the available electrodes voltages and on the CCD dimension sensors. b-c) Ion chain with optimized spacing inhomogeneity for $N=20, 28, 36, 44$ ions. A not-to-scale plot of the quartic potential for N=44 is shown as a guide to the eye. The error bars have been calculated propagating the gaussian fit errors on the ions centers. c) The spacing variance to mean ratio computed numerically for $N=20, 28, 36, 44$ ions as a function of $1/\beta$, which characterizes the ratio between the quadratic and the quartic component.
}
\label{fig_spacing}
\end{figure*}
In order to measure the pressure, we also considered the rate of reconfiguration events $\gamma_{el}$ caused by elastic collisions when $N$ ions are in a zig-zag configuration \cite{Fishman_2008}, namely when $\omega_{y,z}/\omega_x<N/\sqrt{\log(N)},$ where $\omega_{y,z}$ are the two transverse frequencies and $\omega_x$ is the axial frequency (see Fig. \ref{fig_elastic_rate}). When this condition is met, there are two degenerate configurations separated by a small energy gap, which depends on the transverse mode splitting $\Delta\omega_{\rm tr}=\omega_z-\omega_y$. When an elastic collision occurs, the ions have a finite probability to switch from the ``zig'' to the ``zag'' configuration, if the energy gained by the ions is enough to overcome the barrier.  Therefore, our observable is  $\gamma_{\rm el}=p_{\rm flip} \gamma$, where $p_{\rm flip}$ is the probability of flipping the zig-zag chain, which is a function of the transverse mode splitting $\Delta\omega_{\rm tr}$ and of the energy of the incoming particles. In order to calculate $p_{\rm flip}$, we performed a numerical simulation of 31 ions in a Paul trap after a collision with a $H_2$ molecule with mean energy $\langle E_{H_2}\rangle=3/2\, k_B  T$ over $2\times 10^4$ rf periods, calculating the probability dependence on the temperature and on the energy barrier (for details, see \ref{app_numerical_simulation}). The elastic rate $\gamma_{\rm el}$ has been measured  at $\Delta\omega_{\rm tr}=2\pi\times 2~\mbox{kHz}$ at two different temperatures ($T = 4.7$ K and the $T=7$ K). By inverting Eq. (\ref{eq:Pressure}) and using the calculated $p_{\rm flip}$ values, the pressure has been measured to be $P= (2 \pm 1) \cdot 10^{-12}~\mbox{Torr}$ and $P= (4 \pm 2) \cdot 10^{-12}~\mbox{Torr}$ for $T=4.7$ K and 7 K, respectively. 

\section{Uniformly spaced ion-chains}\label{uniform}

The segmented blade trap features 10 static electrodes which allow for the realization of axial anharmonic potentials. In particular, quartic axial potentials have been shown to offer several advantages in handling, cooling \cite{Lin_2015} and performing coherent operations \cite{Lin_2009} on large ion chains. Such potentials relax the conditions necessary to achieve a linear configuration of the ion chain, avoiding zig-zag transitions at the center of the chain. At the same time, tailored anharmonicity in the axial potential allows us to minimize the ion spacing inhomogeneity induced by the Coulomb repulsion and enables control of the average spacing between the ions. In addition, a homogeneous spacing configuration prevents the ions in the chain center from coming too close to each other, which reduces cross-talk in ion state detection and single ion manipulation with focused laser beams. Finally, the anharmonic potentials can be used to shape the normal mode structure \cite{Home2011} and to minimize the inhomogeneity in the laser-induced spin-spin coupling \cite{porras04} by shaping the transverse normal-mode dispersion.

The form of the quartic axial potential induced by the static electrodes can be written as:
\begin{equation}
\label{eq:}
V_{ax}=\sum_{i=1}^N \frac{\alpha_2}{2} x_i^2+\frac{\alpha_4}{4} x_i^4,
\end{equation}
where $x_i$ is the position of the $i$-th ion and $\alpha_2=m \omega_x^2$, with $\omega_x$ the axial frequency and $m$ the ion atomic mass.
The axial equilibrium positions are determined by the balance between the electrostatic axial forces $(F_{ax}=- \partial_x V_{ax})$ and the inter-ion Coulomb repulsion:
\begin{equation}
\label{eq_position}
0=u_i + \beta u_i^3 - \sum_{j=1}^{i-1}(u_i - u_j)^{-2} + \sum_{j=i+1}^{N}(u_i - u_j)^{-2}
\end{equation}
where $u_i=x_i/\ell$ is the $i$-th ion's position in adimensional units and $\ell=(q^2/4\pi \epsilon_0 m\omega_x^2)^{1/3}$ is the characteristic length of the axial potential, with $\epsilon_0$ the vacuum permittivity and $q$ the electron charge. The quartic potential is characterized by the dimensionless ratio $\beta=\alpha_4 \ell^2/\alpha_2$ \cite{Lin_2009}, which can be optimized to minimize the inhomogeneity of the ion chain. By solving equation (\ref{eq_position}), it is possible to find the optimal $\beta^*(N)$ (see Fig. \ref{fig_spacing}c) for a certain ion number $N$, that minimizes the ion inhomogeneity which can be parameterized by the ratio $\sigma_{\Delta x}/\overline{\Delta x}$, where $\sigma_{\Delta x}$ and $\overline{\Delta x}$ are the ion spacing standard deviation and mean spacing, respectively. Once the optimal $\beta^*(N)$ is found, it is possible to fully determine the potential by choosing a desired average spacing $\overline{\Delta x}$, which depends on the absolute value of the harmonic frequency $\omega_x$. A particular choice of the quadratic term $\alpha_2^*$ determines the characteristic length $\ell(\alpha_2^*)$ and corresponds to a certain equilibrium average spacing $\overline{\Delta x}^*$. Therefore, the quartic component is univocally determined by the equation $\alpha_4=\beta^* \alpha_2^*/\ell(\alpha_2^*)$. In Fig. \ref{fig_spacing}a, it is shown how we can obtain the minimum variance to mean spacing ratio (solid curves) by tuning the quartic potential, ranging from $N=18$ to $N=44$  ions in a linear configuration.

\section{Conclusions}\label{sec_conclusions}
In this work we have reported on a new cryogenic ion trap experimental apparatus designed for large scale quantum simulation of spin models. We designed and optimized the system to routinely trap over 100 ions in a linear chain and hold them for hours. We improved the system's mechanical stability, characterized the residual background pressure by measuring the inelastic and elastic collision rate with H$_2$ molecules and used anharmonic potentials to achieve uniformly spaced ion chains. In the short term we plan to address the ions with global Raman beams, inducing tunable spin-spin interaction with power law range \cite{porras04} and to study the dynamics of the system after a quantum quench. In order to compute the exact dynamics in this diabatic regime it is necessary to consider a large fraction of the Hilbert space, which grows exponentially with the system size. This apparatus is also well-suited for implementing single ion addressability using individual light shifts from a laser focused on the ions \cite{Lee2016} through the high numerical aperture window already used for imaging. This tool can be used to realize disordered potentials on the ions or to prepare arbitrary initial product states. In conclusion, this new system will be an ideal test-bed to investigate quantum quenches with a variety of spin 1/2 Hamiltonians with Ising or XY interactions and with a number of qubits impossible to simulate exactly on a classical computer.

\section*{ Acknowledgements}
We would like to acknowledge K. Beck, M. Cetina, L.-M. Duan, R. Blumel, P. Julienne, B. Ruzic, and Y. Nam for useful discussions; D. Stick and A. Casias at Sandia National Laboratories for ion trap fabrication; and A. Restelli for technical assistance. This work is supported by the ARO and AFOSR Atomic and Molecular Physics Programs, the AFOSR MURI on Quantum Measurement and Verification, the IARPA LogiQ program, the ARO MURI on Modular Quantum Systems, the ARL Center for Distributed Quantum Information, the IC Postdoctoral Fellowship Program, and the NSF Physics Frontier Center at JQI.

\clearpage
\section*{References}
\bibliographystyle{iopart-num}
\bibliography{Cryolibrary_iop_1}

\clearpage
\appendix

\section{Heat load budget}\label{app_heat_loads}
\begin{center}
\label{table_heat_loads}
\begin{tabular}{|c|c|c|}
\hline
Component & $\dot{Q}_{40 K}~\mbox{[W]}$ & $\dot{Q}_{4 K}~\mbox{[mW]}$  \\
\hline
40 K shield & 5.6 & -  \\
4 K shield & - & 1 \\
2'' viewport & 0.8 & 0.6 \\
1'' viewports & 1.6 & 1.2 \\
wiring &  0.5 &  220 \\
\hline
\end{tabular}
\end{center}

\section{Ion-Molecule Collision}\label{app:collision}
Let us consider the $1/r^4$ polarization interaction between the ion and the neutral molecule. Suppose the reduced mass is $\mu=m_1 m_2/(m_1+m_2)$, angular momentum $L$ and energy $E$ in the center of mass frame, and the attractive potential $V(r)=-C_4/ r^4$, where $C_4= \alpha e^2/8\pi\varepsilon_0$. They are related to the initial velocity $v_0$ and the impact parameter $b$ in the lab frame as $L=b\mu v_0$ and $E=\frac{1}{2}\mu v_0^2$. We can use the Binet equation to solve the shape of the orbit  \cite{Langevin1905}:
\begin{equation}
\theta = \int \frac{L du}{\sqrt{2\mu (E + C_4 u^4)-L^2 u^2}},
\end{equation}
where $u\equiv 1/r$. Here we are interested in the scattering angle of the particles:
\begin{equation}
\pi - \theta_s = 2\int_0^{u_0} \frac{L du}{\sqrt{2\mu (E + C_4 u^4)-L^2 u^2}},
\end{equation}
where
\begin{equation}
u_0 = \sqrt{\frac{L^2-\sqrt{L^4 - 16 C_4 \mu^2 E}}{\alpha\mu}}
\end{equation}
corresponds to the inverse minimal distance between the two particles. After some algebra we get
\begin{equation}
\theta_{\rm sc} = \pi - \frac{2\sqrt{2}K\left(\frac{1-x}{1+x}\right)}{\sqrt{1+x}},
\end{equation}
where $x=\sqrt{1-\frac{16 C_4 \mu^2 E}{L^4}}$ and $K(m)$ is the complete elliptic integral of the first kind. In the limit $C_4\to 0$ we have $\theta_s\to 0$. Note that this expression works only if $16 C_4 \mu^2 E<L^4$, otherwise the two particles will really collide as they approach each other. This condition defines a critical impact parameter $b_c=(8 C_4 /\mu v_0^2)^{1/4}$. Finally, the velocity of the ion in the lab frame, which is originally 0, becomes
\begin{eqnarray}
v_x &=& \frac{M_{\mathrm{m}}}{M_{\mathrm{i}}+M_{\mathrm{m}}} v_0 (1-\cos \theta_{\rm sc}),\\
v_y &=& -\frac{M_{\mathrm{m}}}{M_{\mathrm{i}}+M_{\mathrm{m}}} v_0 \sin \theta_{\rm sc}.
\end{eqnarray}

In the $16 C_4 \mu^2 E>L^4$ case, as mentioned above, the two particles will keep being attracted to each other. The rotation angle when they overlap is
\begin{equation}
\theta = \int_0^{\infty} \frac{dx}{\sqrt{1-x^2+\frac{4 C_4\mu^2 E}{L^4}x^4}},
\end{equation}
leading to $\theta_{\rm sc}=\pi-2\theta$. For the following numerical simulation, we considered $\theta_{\rm sc}=\pi$ for simplicity. 
This will overestimate by a factor of order one the energy transferred to the ion for $b<b_c$. However as we truncate the impact parameter at $b_c$, we are also underestimating the energy transfer for $b>b_c$. Therefore the overall effect of this simplification should not be significant. For a more rigorous treatment, it would be required to integrate over the impact parameter to get the average amount of energy transfer in the collision, or sample directly the impact parameter for each simulation to find the energy of the ion after the collision.

\section{Numerical Simulation of a zig-zag Ion Chain}\label{app_numerical_simulation}
\begin{figure}[htbp]
  \centering
  \includegraphics[width=0.5\linewidth]{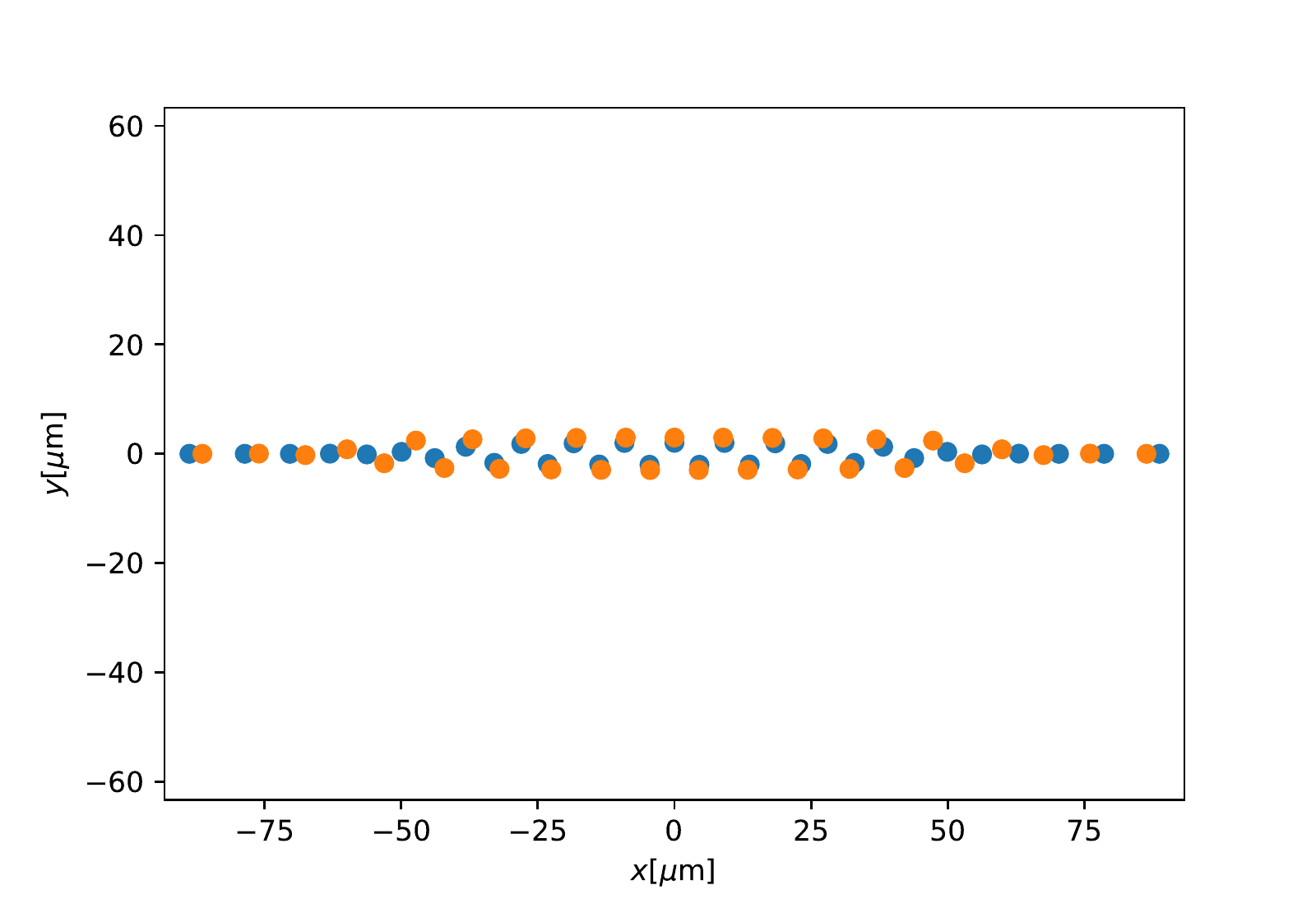}\\
  \caption{Equilibrium positions of ions (at the beginning of an rf cycle) for $\Delta\omega_{\rm tr}=2\pi\times1\,$kHz (blue) and $\Delta\omega_{\rm tr}=2\pi\times299\,$kHz (orange).}\label{fig:zigzag}
\end{figure}

Let us consider a trap with $\Omega_{\rm rf}=2\pi\times 24\,$MHz and $\omega_x=2\pi\times 67\,$kHz, $(\omega_y+\omega_z)/2=2\pi\times 622.5\,$kHz. For $N=31$ ions and transverse splitting up to $\Delta\omega_{\rm tr}/2\pi = (\omega_z-\omega_y)/2\pi=299\,$kHz, the equilibrium positions of the ions have a zig-zag shape in the $x-y$ plane. For example, Fig.~\ref{fig:zigzag} shows the equilibrium positions for $\Delta\omega_{\rm tr}=2\pi\times1\,$kHz (blue) and $\Delta\omega_{\rm tr}=2\pi\times299\,$kHz (orange).

For a given $\Delta\omega_{\rm tr}$, hence given trap parameters and equilibrium positions, we can simulate the probability for the zig-zag shape to flip after a collision with a $\mathrm{H}_2$ molecule at temperature $T$. The velocity of $\mathrm{H}_2$ follows a Maxwell-Boltzmann distribution and, as mentioned above, the collision is in 1D along the incoming direction of the molecule. Numerically, we find that after $2\times 10^4$ rf periods the ions will reach a local potential minimum (not necessarily the original one) due to the Doppler cooling effect (see also \ref{app:Doppler}). Then, we can compare the initial and the final configurations and see if the chain has flipped from a ``zig'' to a ``zag'' configuration. At low temperature such probability is very small and hence lots of samples are needed to measure it precisely. In order to count the number of flipping events, we sort the initial and the final positions respectively according to their $x$ coordinates, and then add up the squared distances between corresponding ions. If the configuration has not flipped, the sum will be small (in our case $\lesssim 22 \,\mu{\rm m}^2$); otherwise the sum will be much larger. 

From the computational point of view, it has to be pointed out that the trapped ion system is chaotic because of the nonlinear Coulomb interaction, thus any computational error will accumulate exponentially with the evolution time. To suppress the error, we may need very small step length and very high numerical precision, and the simulation soon becomes computationally intractable. However, we know that in the experiment there is always a Doppler cooling beam (see ~\ref{app:Doppler} for how this effect is included in the simulation) and that the scattering of the photons itself is random. Therefore small errors in the simulation shall not change the qualitative behavior. For the simulations we use Forest-Ruth method with 100 steps per rf period and \emph{double} precision. We checked $p_{\rm flip}$ convergence increasing from 100 to 1000 steps per RF period.

\begin{figure}[htbp]
  \centering
  \includegraphics[width=0.6\columnwidth]{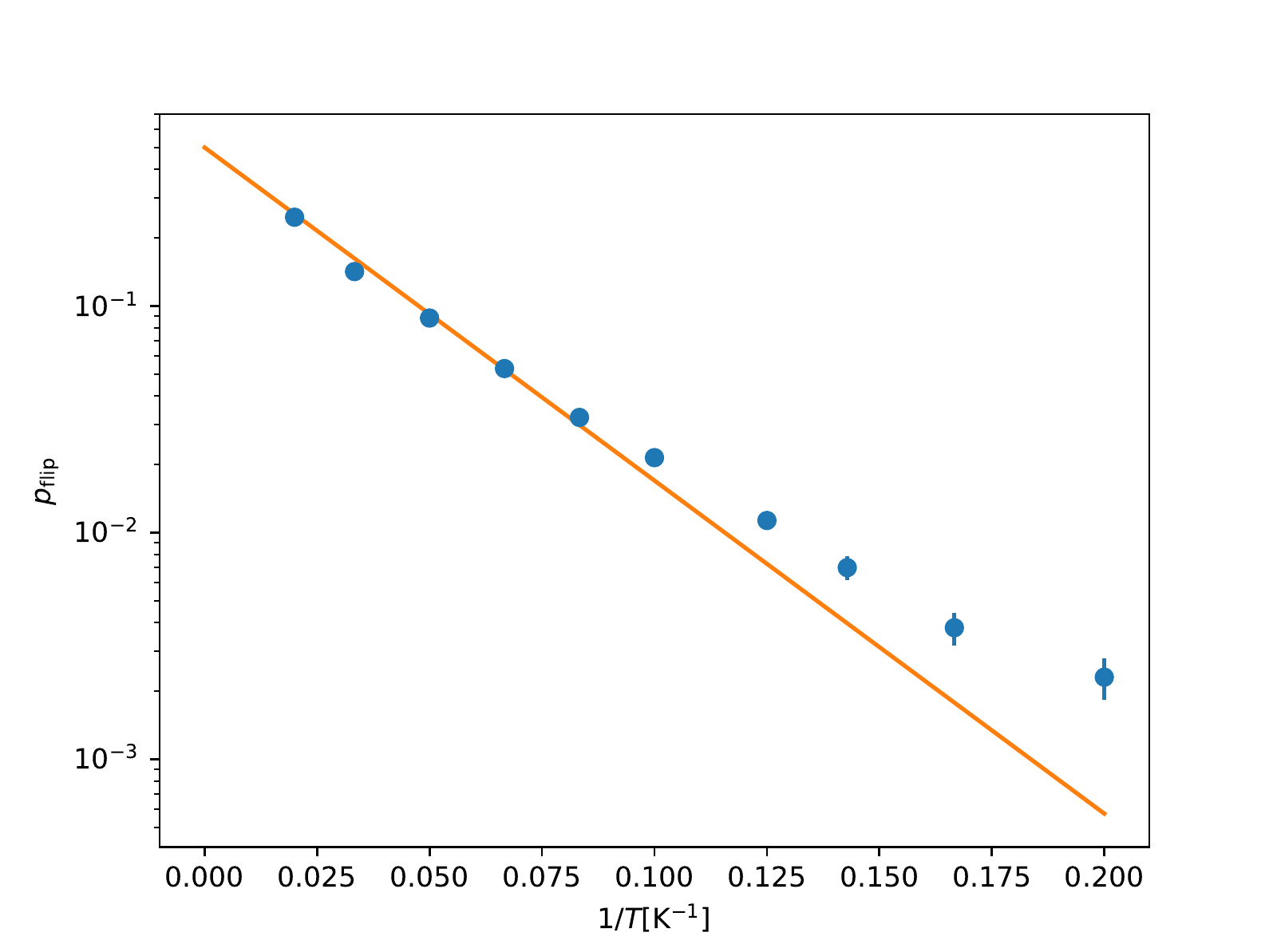}\\
  \caption{$p_{\rm flip}$ vs. $T$ for $\Delta\omega_{\rm tr}=2\pi\times 19\,$kHz.}\label{fig:PvsT}
\end{figure}

First, we study the $T$ dependence in the probability $p_{\rm flip}$ by fixing $\Delta\omega_{\rm tr}=2\pi\times19\,$kHz. The numerical results are shown in Fig.~\ref{fig:PvsT}.
As we can see, for high temperature ($T\gtrsim 12\,$K) the probability can be modelled as $p_{\rm flip}=0.5\exp(-E/T)$; but at low temperature the probability is greater than this prediction. This suggests the existence of multiple paths and hence multiple energy barriers to flip the configuration.

\begin{figure}[htbp]
  \centering
  \includegraphics[width=1\columnwidth]{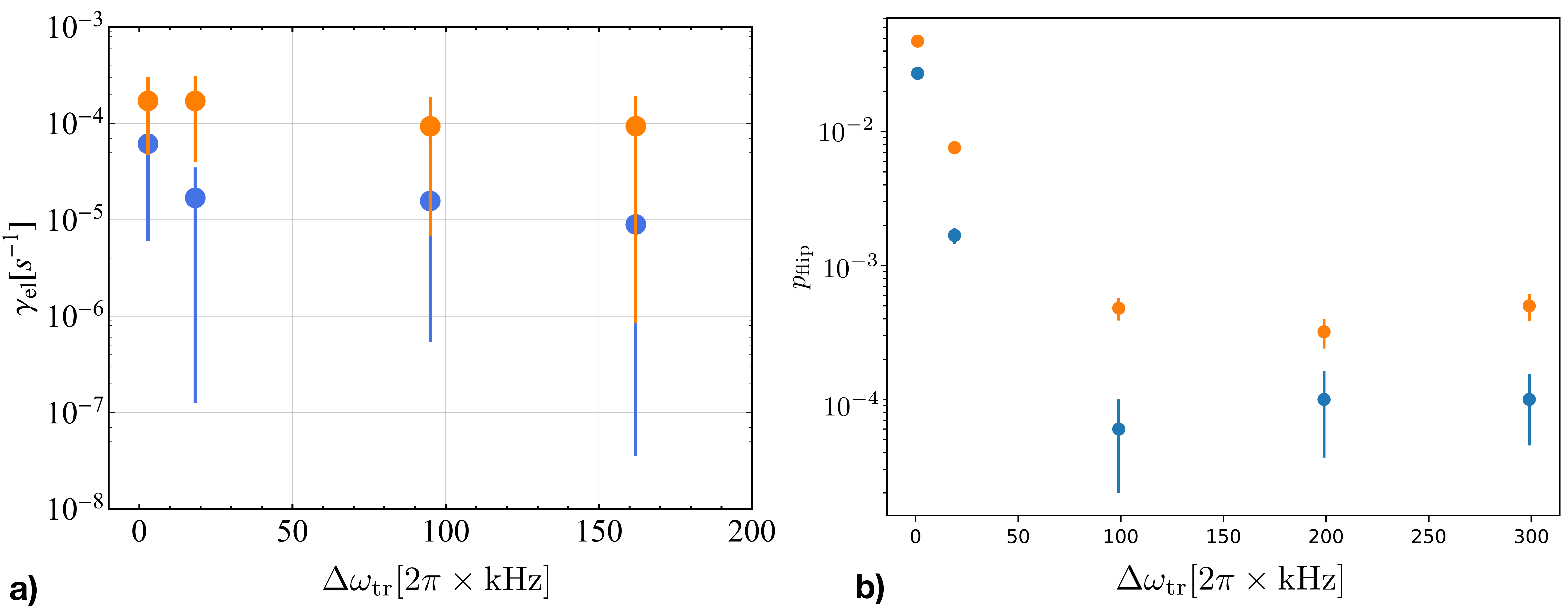}\\
  \caption{ {\bf a)} $ \gamma_{el}$ as a function of the barrier $\Delta\omega_{\rm tr}={\omega_z-\omega_y}$. The blue (orange) points refer to $T= 4.5\, {\rm K}\, (T= 7 \,{\rm K})$, respectively. The data have been acquired with a number of ions varying from $N=31$ to $N=38$ for a time period varying up to 12 hours. The reported rate is per ion. {\bf b)} Numerical results for $p_{\rm flip}$ vs. $\Delta\omega_{\rm tr}$ for  $T=4.7\,$K (blue) and $T=7\,$K (orange). 
}\label{fig:PvsDelta}
\end{figure}

Next, we consider the $\Delta\omega_{\rm tr}$ dependence at $T=4.7\,$K and $T=7\,$K, which is shown in Fig.~\ref{fig:PvsDelta}b. 
Because the probability $p_{\mathrm{flip}}$ is so low, we randomly sample $10^5$ initial conditions to estimate the probability and repeat this process for 5 times to estimate the error bar. Note that $p_{\rm flip}$ does not decrease monotonically as $\Delta\omega_{\rm tr}$ increases. This is because of the multiple modes inside the $y$-$z$ plane, i.e. the previously mentioned multiple energy barriers. 
Also note that at large $\Delta\omega_{\rm tr}$ it is possible to flip ``half'' the configuration: the ions will reach a local minimum of potential energy, but not necessarily the global minimum.

Fig. \ref{fig:PvsDelta}a shows the elastic rate measurements as a function of the barrier $\Delta\omega_{\rm tr}$.  As the barrier increases, there is a suppression in the reconfiguration rate but not of two orders of magnitude as predicted by the theory (Fig. \ref{fig:PvsDelta}b). The discrepancy might be due to the difficulty in acquiring data for hours and in sampling such a low reconfiguration probability without having flipping events not caused by collisions. We experimentally observed that the flipping rate increases with the number of dark ions, which are only sympathetically cooled by the bright ions, leading to a higher temperature of the chain that can cause the thermal activation of the flipping events. Also electric field noise (not included in the simulation) at frequencies of the order of the barrier could flip the zig zag chain. These two factors probably prevent us from resolving reconfiguration rates lower than $10^{-5} ~\mbox{s$^{-1}$}$. For these reasons the measurement of $\gamma_{\rm el}$ at the lowest barrier value $\Delta\omega_{\rm tr}/(2\pi)=2~\mbox{kHz}$ is the most reliable estimate of the pressure.

\section{Doppler Cooling}
\label{app:Doppler}
A weak Doppler cooling beam is always on during the storage of the ion crystal. Typically, the Doppler Cooling beam at $\lambda=369~\mbox{nm}$ is directed along the $[1,1,0]$ direction of the trap and detuned at $\Delta=-\Gamma/2$ ($\Gamma_{^{2}P_{1/2}}= 2\pi\times19~\mbox{MHz}$ in $^{171}$Yb$^+$) with a saturation parameter $s=I/I_s=1$, where $I_s=4\pi^2\hbar c\Gamma/6\lambda^3$ is the saturation intensity. Since with these parameters the photon scattering rate is about $2\pi\times3$ MHz and is smaller than the rf frequency of the trap, the damping force should not be modelled as a simple $-\gamma {\bf v}$ term. To describe the physics here we use $N$ two-dimensional vectors to describe the internal states of the ions and consider the excitation of ions by the laser beam quantum mechanically. Then, quantum trajectory \cite{Andrew2014} method can be used to simulate the scattering process. At each step the internal state of an ion is first evolved from $|\psi(t)\rangle$ to $|\tilde{\psi}(t+dt)\rangle$ under the effective Hamiltonian
\begin{equation}
H_{\mathrm{eff}} = \frac{\Delta-{\bf k}\cdot {\bf v}}{2}\sigma_z + \frac{\Omega}{2} \sigma_x - \frac{i}{4}\gamma(I-\sigma_z).
\end{equation}

Then, with probability $\langle \tilde{\psi}(t+dt)| \tilde{\psi}(t+dt) \rangle$ the internal state is renormalized to $|\psi(t+dt)\rangle=|\tilde{\psi}(t+dt)\rangle/\||\tilde{\psi}(t+dt)\rangle\|$; otherwise a transition occurs and the ion goes back to the ground state $|\psi(t+dt)\rangle=|0\rangle$, together with a momentum change as the ion absorbs a photon in the $[1,1,0]$ direction and then emits one into a random direction.


\end{document}